\documentclass[a4paper,12pt]{article}

\usepackage{amsmath}
\usepackage{latexsym}
\usepackage{cite}
\usepackage{graphicx}

\newcommand{\fdual}{\tilde{F}}

\newcommand{\first}{\ensuremath{\mbox{1}^{\mbox{\tiny st}}}}

\newcommand{\lomega}{\bra{\Omega}}
\newcommand{\lrcovar}{{\stackrel{\leftrightarrow}{D}\!}\mbox{\hspace{0.17ex}}}

\newcommand{\lrpartial}{{\stackrel{\leftrightarrow}{\partial}\!}\mbox{\hspace{0.17ex}}}

\newcommand{\medsp}{\\[0.7ex]}
\newcommand{\spsp}{\\[0.35ex]}

\newcommand{\romega}{\ket{\Omega}}
\newcommand{\second}{\ensuremath{\mbox{2}^{\mbox{\tiny nd}}}}

\newcommand{\suppot}{\mathcal{W}}

\newcommand{\unj}{\underline{j}}
\newcommand{\unk}{\underline{k}}
\newcommand{\unl}{\underline{l}}
\newcommand{\uone}{\mbox{\slshape U}(1)}

\newcommand{\Ltext}[1]{\ensuremath{\itindex{\mathcal{L}}{#1}}}

\newcommand{\bra}[1]{\langle #1 |}

\newcommand{\diff}[1]{\mbox{d}#1}

\newcommand{\half}[1]{\ensuremath{\frac{#1}{2}}}
\newcommand{\intd}[1]{\int \!\! #1 \;}
\newcommand{\inv}[1]{\ensuremath{\frac{1}{#1}}}
\newcommand{\ket}[1]{| #1 \rangle}
\newcommand{\metr}[1][]{g_{\varphi \bar{\varphi} #1}}

\newcommand{\wphi}[1][]{\suppot_{,\varphi #1}}
\newcommand{\bwphi}[1][]{\bar{\suppot}_{,\bar{\varphi} #1}}

\newcommand{\itindex}[2]{\ensuremath{#1_{\mbox{\scriptsize{\itshape #2}}}}}

\DeclareMathOperator{\hc}{h.c.}

\begin{document}
\bibliographystyle{phaip}

\renewcommand{\thefootnote}{\fnsymbol{footnote}}
\thispagestyle{empty}
\begin{titlepage}

\vspace*{-1.5cm}
\hfill \parbox{3.5cm}{hep-th/0205240}
\vspace*{1.0cm}

\begin{center}
  {\Large {\bf \hspace*{-0.2cm} On the Effective Description
      \protect\vspace*{0.2cm} \\
      \protect\hspace*{0.2cm} of Dynamically Broken
      \protect\vspace*{0.4cm} \\      
      \protect\hspace*{0.2cm} SUSY-Theories}}
  \vspace*{1.5cm} \\

{\bf L. Bergamin\footnote{email: bergamin@tph.tuwien.ac.at, phone: +43  1 58801 13622, fax: +43 1 58801 13699}}\\
      Institute for Theoretical Physics \\
    Technical University of Vienna \\
    Wiedner Hauptstr.\ 8-10 \\
    A-1040 Vienna, Austria\\[3ex]
{\bf P. Minkowski\footnote{email: mink@itp.unibe.ch, phone: +41 31 631 8624,
    fax: +41 31 631 3821}}\\
    Institute for Theoretical Physics \\
    University of Bern \\
    Sidlerstrasse 5\\
    CH - 3012 Bern, Switzerland
   \vspace*{0.8cm} \\  


\vspace*{1.3cm}

\begin{abstract}
\noindent
We introduce a new class of effective actions describing dynamically broken
supersymmetric theories in an essentially non-perturbative region. Our approach is a
generalization of the known supersymmetric non-linear sigma models, but allows
in contrast to the latter the description of dynamical supersymmetry breaking
by non-perturbative non-semiclassical effects. This non-perturbative breaking
mechanism takes place in confined theories, where the effective fields are
composite operators. It is necessary within the context of quantum
effective actions and the associated concept of symmetry breaking as a
hysteresis effect. In this paper we provide a mathematical definition and
description of the actions, its application to specific supersymmetic gauge
theories is presented elsewhere.

\vspace{3mm}
\noindent
{\footnotesize {\it PACS:} 11.30.Qc; 11.30.Pb; 11.15.Kc \newline
{\it keywords:} dynamical supersymmetry breaking; low
energy approximations; SYM theory}
\end{abstract}
\end{center}

\end{titlepage}

\renewcommand{\thefootnote}{\arabic{footnote}}
\setcounter{footnote}{0}
\numberwithin{equation}{section}
%
\section{Introduction and Outline of the Problem}
\label{sec:idea}
Effective Lagrangian techniques for about twenty years have been a successful
tool to explore the low-energy dynamics of supersymmetric
gauge-theories. While most of the models investigated are in the Higgs or
Coulomb phase at low energies, mainly due to the work by Dijkgraaf and Vafa
\cite{Dijkgraaf:2002fc,Dijkgraaf:2002vw,Dijkgraaf:2002dh,Dijkgraaf:2002xd}
confined models are of particular interest in these days. There exists a
simple but far-reaching difference between confined theories and theories in
the Higgs or Coulomb phase: The effective fields of a confined theory are
always composite operators. Although this is not solely a technical
difficulty, but a main characteristic of the phenomenology of confined
theories, a detailed discussion of its consequences for the construction of effective actions within
supersymmetry does not seem to exist in the literature. In this paper we
propose a consistent treatment of composite effective fields within
supersymmetry and we show that confinement indeed leads to fundamental changes
in the understanding of the effective action.

Before going into the technical details of our construction we shortly review
the problems arising in the context of composite effective superfields. As a
standard example of a confined supersymmetric theory we consider $N=1$ SYM. We want to construct an effective action for $N=1$ SYM theories based on
classical fields from composite operators that represent (by assumption) the
relevant low energy degrees of freedom. It has been shown that we obtain such
an effective action by extending the complex coupling constant $\tau$ of SYM
to a chiral superfield $J(x) = \tau(x) + \theta \eta - 2 \theta^2 m$
\cite{veneziano82, bib:mamink, bib:markus}. The effective action is obtained
by Legendre transformation and is formulated in terms of three classical
fields $\varphi$, $\psi$ and $F$ that represent the gluino condensate, a
spinor (the goldstino in case of dynamical supersymmetry breaking) and the classical Lagrangian, respectively. The source extension is
unique in the sense that there exists no other extension that preserves gauge
invariance and supersymmetry covariance \cite{bergamin01:2}. By combining the
three effective fields to a chiral superfield $\Phi = \varphi + \theta \psi +
\theta^2 F$, the effective action can be written as an integral over superspace.

This system had been studied in refs.\ \cite{veneziano82,burgess95,bib:mamink}, the
ansatz for the effective action used therein can be written as
\begin{equation}
\label{eq:geomeffact}
  \mathcal{L} = \intd{\diff{^4x}} \biggl( \intd{\diff{^4 \theta}} K(\Phi,
  \bar{\Phi}) - \bigl(\intd{\diff{^2 \theta}} W(\Phi) + \hc \bigr) \biggr)\ ,
\end{equation}
where the superpotential is determined by the anomaly-structure
\begin{equation}
  W(\Phi) = \Phi (\log \frac{z \Phi}{\Lambda^3} - 1)
\end{equation}
and the K\"{a}hler potential is a \emph{polynomial} function in $\Phi$. The
most important consequence from the ansatz \eqref{eq:geomeffact} is: \emph{The
effective composite operator $F$ must be an auxiliary field.} As this effective
field is build up from composite operators, this constraint is neither
motivated nor consistent. Indeed, in this situation there exist two types of
``auxiliary'' fields:
\begin{enumerate}
\item The auxiliary field of the fundamental theory, in $N = 1$ SYM typically
  denoted by $D$. The theory must be ultra-local in this field and it can be
  eliminated consistently even in the full quantum theory. This field will be
  denoted as \first\ generation auxiliary field in the following.
\item The effective composite field appearing in the effective superfield at a
  place, where one usually expects an auxiliary field ($F$ in $N=1$ SYM, \second\
  generation ``auxiliary'' field in the following). As the field $F$ is itself
  a composite operator of fundamental fields, it is not directly related to
  the \first\ generation auxiliary field $D$. It even exists, if the \first\
  generation field has been eliminated by its algebraic equations of motion!
\end{enumerate}
The imperative to distinguish strictly between \first\ and \second\
generation ``auxiliary'' fields must lead to the conclusion that the ansatz
\eqref{eq:geomeffact} is inconsistent. This has been discussed extensively in
\cite{bergamin01}, in the following the main conclusions are summarized:
\begin{enumerate}
\item \label{enum1}Within the ansatz \eqref{eq:geomeffact} the field $F$ has the typical
potential of an auxiliary field:  It
is not bounded from below but falls off to $- \infty$ as $F$ goes to $+
\infty$. In fact it does not even have a local minimum, but instead an
absolute maximum, cf.\ figure \ref{fig:potential}. All supersymmetric Lagrangians
with a polynomial K\"{a}hler potential have auxiliary fields with a
potential of this type. This implies that one is forced to interpret the
physical ground-state of the theory by the absolute maximum of the auxiliary
field potential. 
\item \label{enum2} Point \ref{enum1} is equivalent to the statement that $F$ must be an auxiliary field
  that can be eliminated by its algebraic equations of motion. Indeed, the
  spectrum of the theory obtained in \cite{veneziano82} can be found
  \emph{after} this elimination, only.
\item \label{enum3}As a consequence of \ref{enum1} and \ref{enum2} the action must be
  ultra-local in $F$ \emph{exactly}, i.e.\ derivatives acting on $F$ do not
  exist. This point is important as the action \eqref{eq:geomeffact} is
  certainly not the complete effective action of $N=1$ SYM, but at most an
  approximation for small momenta ($p^2 \ll \Lambda^2$). The fact that $F$
  must be auxiliary then means that derivative terms on this field do not
  vanish solely in the approximation \eqref{eq:geomeffact}, but \emph{within
    the complete effective action}. Indeed, when introducing derivative terms on $F$
  as higher order effects  the auxiliary fields become dynamical\footnote{We
    emphasize that we are using a quantum effective
    action, whose description is a purely classical object. Indeed the
    question of dynamical auxiliary fields is quite different, if the
    non-linear model is not seen as a classical but a quantum object. We will
    comment on this below.} (the action is no
longer ultra-local in this field). Consequently the absolute maximum must
become an unstable point of the theory and $F$ moves towards the correct
ground-state of this situation: $\lomega F \romega = \infty$! Of course such
effects are suppressed at low energies compared to the dynamics of
$\varphi$. It is of main importance to notice that this is irrelevant: Simple
classical stability considerations show that any finite contribution to the
dynamics of $F$ is sufficient to generate the instability (this point is
discussed more in detail in \cite{Bergamin:2003ub}). Thus either the higher
order derivatives vanish \emph{exactly} or $\lomega F \romega = \infty$ is the
``correct'' ground-state. In the latter case we would have to conclude, that we did not
identify our low-energy degrees of freedom correctly.
\item As $F$ is not related to the fundamental auxiliary field, an
  interpretation as sketched in point
  \ref{enum3} simply fails to capture the physics of this field. $F$ contains
  the operators $F_{\mu \nu} F^{\mu \nu}$ and $F_{\mu \nu} \fdual^{\mu \nu}$,
  which are dynamical variables of pure YM theory and certainly the same
  applies to SYM. The misinterpretation of the composite field $F$ within the
  ansatz \eqref{eq:geomeffact} (points \ref{enum1}-\ref{enum3}) led to the
  mystery of the missing glue-balls: As $F$ has not been interpreted as an
  independent degree of freedom, the resulting theory is (by assumption)
  confined and has a mass-gap, but nevertheless there appears no glue-ball in
  its spectrum.
\item To escape the conclusion of the last point one could try to include the
  glue-ball without identifying it with the operator $F$
  (this has been suggested in \cite{farrar98}, from our point of view the
  result found therein is not consistent with $N=1$ SYM \cite{bergamin01}). But
  this would not solve the problem, as such an effective action would still be
  strictly local (one derivative for the spinor,
  two derivatives for the gluino condensate). If we identified the low-energy
  degrees of freedom correctly higher order derivatives are indeed suppressed
  but they are not allowed to vanish exactly. An effective action of this type
  is not acceptable as long as the underlying theory is not free.

  Besides this general objection a careful analysis within the supersymmetric
  framework shows that the locality of
  supersymmetric non-linear sigma models is not just a harmless peculiarity but has drastic
  consequences: By writing down the effective action as superspace integral we
  assumed that the invariance under extrinsic supersymmetry (transforming the
  quantum fields as well as the sources) is realized thereby and the generic
  form of \eqref{eq:geomeffact} is correct for any value of the sources (up to
  possible spurion fields, which are irrelevant for the following arguments). By
  decoupling the gluino, \eqref{eq:geomeffact} becomes completely
  non-dynamical. From physical arguments we would expect that this effective
  action breaks down at some value of the gluino mass: As we increase $m$ the
  mass of the lightest gluino state increases as well and for some critical
  value $m_c$
  reaches the scale of the lightest glue-ball. At this point the description
  \eqref{eq:geomeffact} should break down as it does not include all relevant
  low-energy degrees of freedom. This should be seen by the breakdown of the expansion in
  the derivatives. But this breakdown can never take place in
  \eqref{eq:geomeffact} as higher order derivatives are always strictly
  zero. Notice that the alternative indication of a breakdown of the
  description --instabilities in the potential-- is excluded as well. For
  details we refer the reader to \cite{bergamin01}.
\end{enumerate}
An acceptable ansatz for the quantum effective action could be found by
  dropping the assumption that the latter can be written as an integral over
  superspace. Without further specifications such a model has been
  discussed in \cite{bergamin01}. There indeed exist possibilities for
  descriptions of this type for both, broken and unbroken supersymmetry. However in its most general form such an ansatz seems to be excluded by
  symmetry-arguments: As the three classical fields do transform under
  supersymmetry they build a representation thereof and
  thus supersymmetry would be realized non-linearly. But this contradicts
  the assumption that we can expand our effective action in the momenta, as
  non-linear representations mix different orders in $p^2$. This would then
  lead to the conclusion that we did not correctly identify the low-energy
  degrees of freedom.

In this paper we show how to construct a model obeying all requirements
outlined above and leading to an effective action for SYM with
\begin{itemize}
\item linear realization of supersymmetry,
\item dynamical glue-ball,
\item infinite orders of derivatives, where higher orders are suppressed but present.
\end{itemize}
As most important consequence of a consistent implementation of these three
points we will find that supersymmetry breaks dynamically. The breaking
mechanism is of essentially non-perturbative character and is not comparable
to any other breaking mechanism known in literature.

In the present paper we explain the mathematical definition the model and its basic
physical properties. The application of these ideas to SYM is worked out in
\cite{Bergamin:2003ub}.


\section{Definition of the Model}
\subsection{The Basic Idea}
In principle a non-local effective Lagrangian and dynamical auxiliary fields do not stand in contradiction
to (linearly realized) supersymmetry. When considering a single chiral superfield $\Phi = \varphi +
\theta \psi + \theta^2 F$, then the expression
\begin{equation}
  \label{eq:unstable}
  \Ltext{kin} = \intd{\diff{^4 \theta}} (c_0 \bar{\Phi} \Phi - c_1 \bar{\Phi} \Box
  \Phi - c_2 \bar{\Phi} \Box^2 \Phi - \ldots )
\end{equation}
is invariant under supersymmetry. As outlined in the introduction the simple
extension of the ansatz \eqref{eq:geomeffact} with such higher order
derivative terms is excluded by stability
arguments as they introduce kinetic terms for
the auxiliary field $F$:
\begin{equation}
  \Ltext{kin} = c_0 |F|^2 - c_1 \bar{F} \Box F - c_2 \bar{F} \Box^2 F + \ldots
\end{equation}
\begin{figure}[t]
\begin{center}
    \includegraphics[{scale=0.5}]{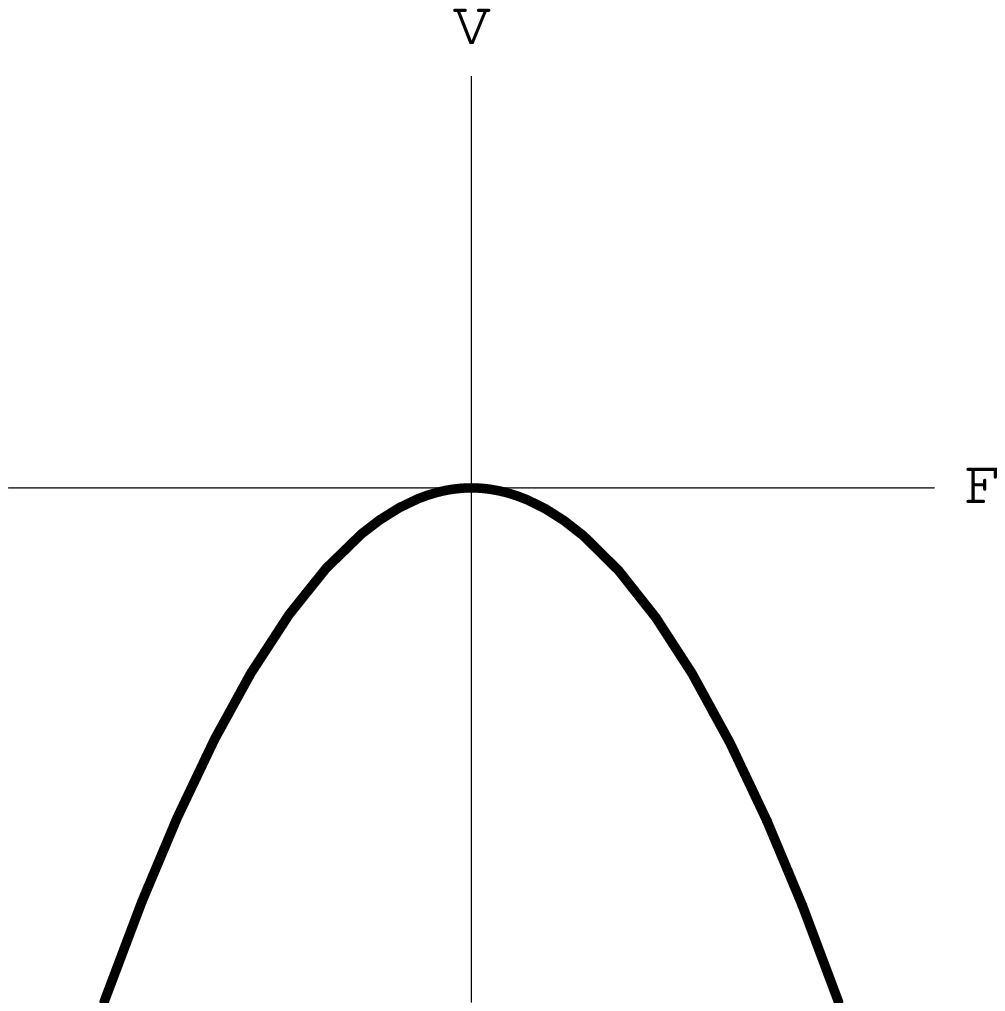} \hspace{3cm}
    \includegraphics[{scale=0.5}]{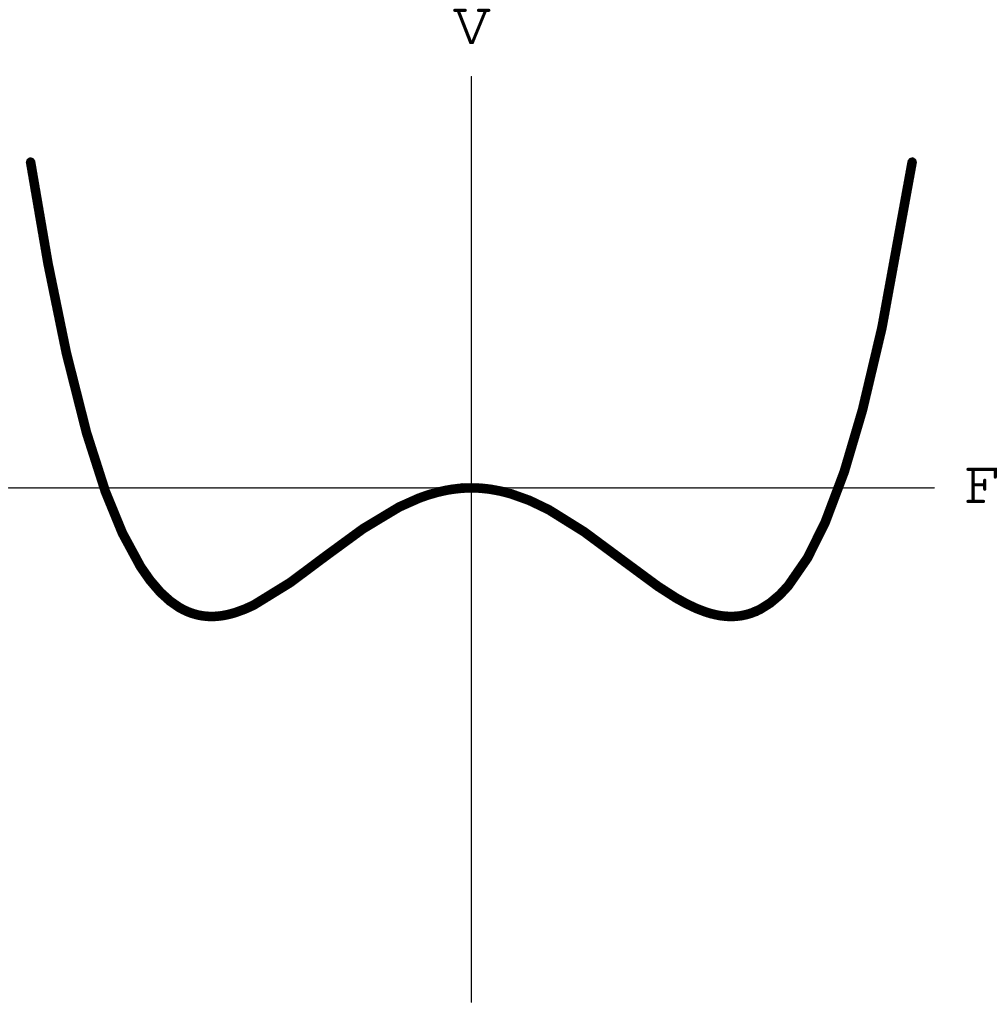}
\end{center}
    \caption{The potential of the highest component $F$ in a theory with
    standard K\"{a}hler potential as non-holomorphic part (left hand side) and
    a possible shape of the potential within the extension proposed in this
    work (right hand side).}
    \label{fig:potential}
\end{figure}
 
To construct a physically meaningful Lagrangian containing terms of the form
\eqref{eq:unstable} it is indispensable to construct a potential for the
\second\ generation ``auxiliary'' field $F$, which is bounded from
below\footnote{In a situation with more than one effective superfield, an
  alternative route to include higher derivative terms has been discussed in
  the literature \cite{nemeschansky85,bergshoeff85,karlhede87,gates97}. But
  within that approach, the \second\ generation auxiliary fields keep their
  usual behavior and thus these models deal with a physically different
  situation than the one discussed here.} (cf.\
figure \ref{fig:potential}). At
the same time of course, the potential in $\varphi$ and $\psi$ must be bounded
from below as well and the dynamics in all fields must be stable at least for
small momenta.

\begin{figure}[t]
\begin{center}
    \includegraphics[{scale=0.3}]{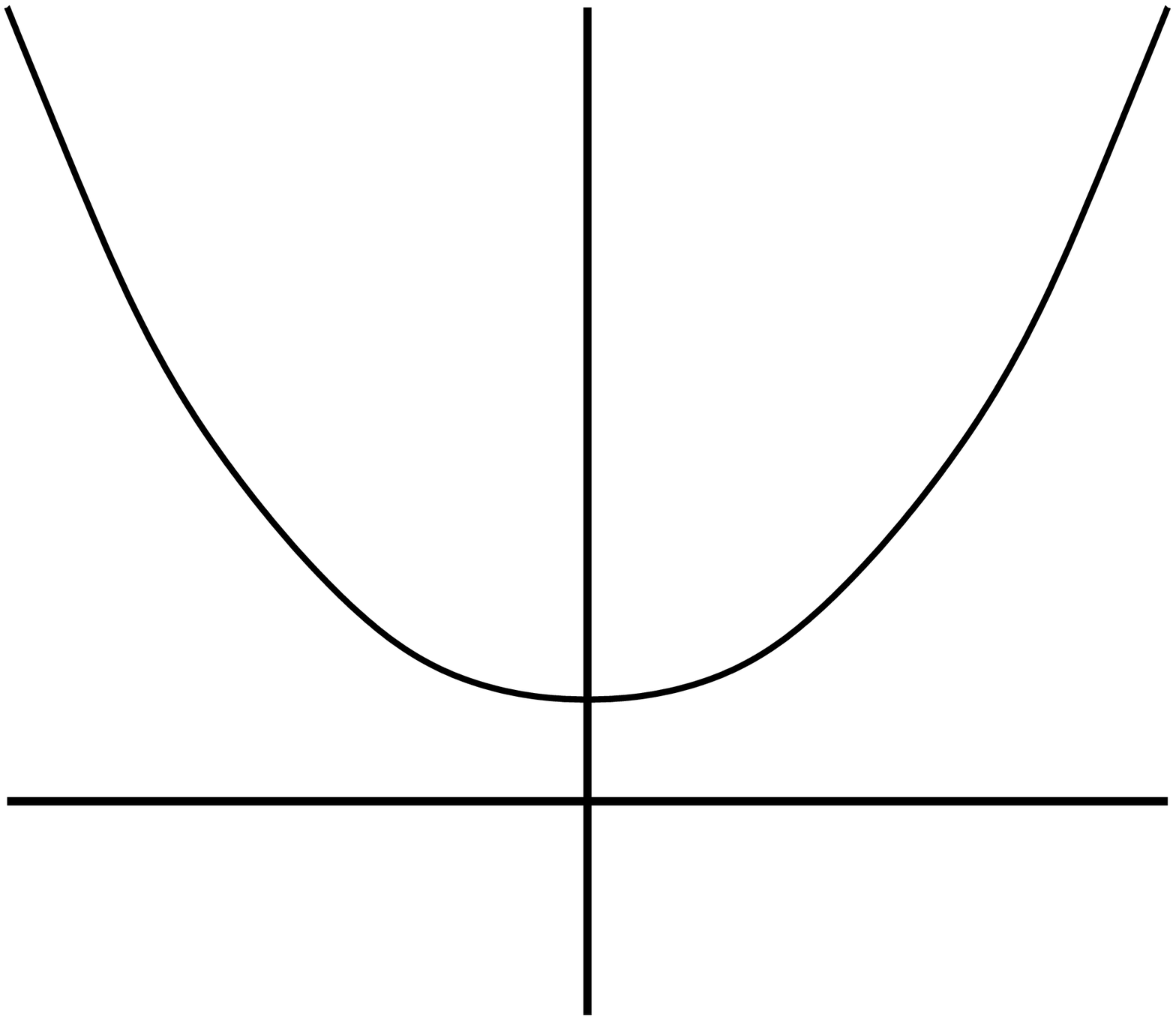} \hspace{3cm}
    \includegraphics[{scale=0.3}]{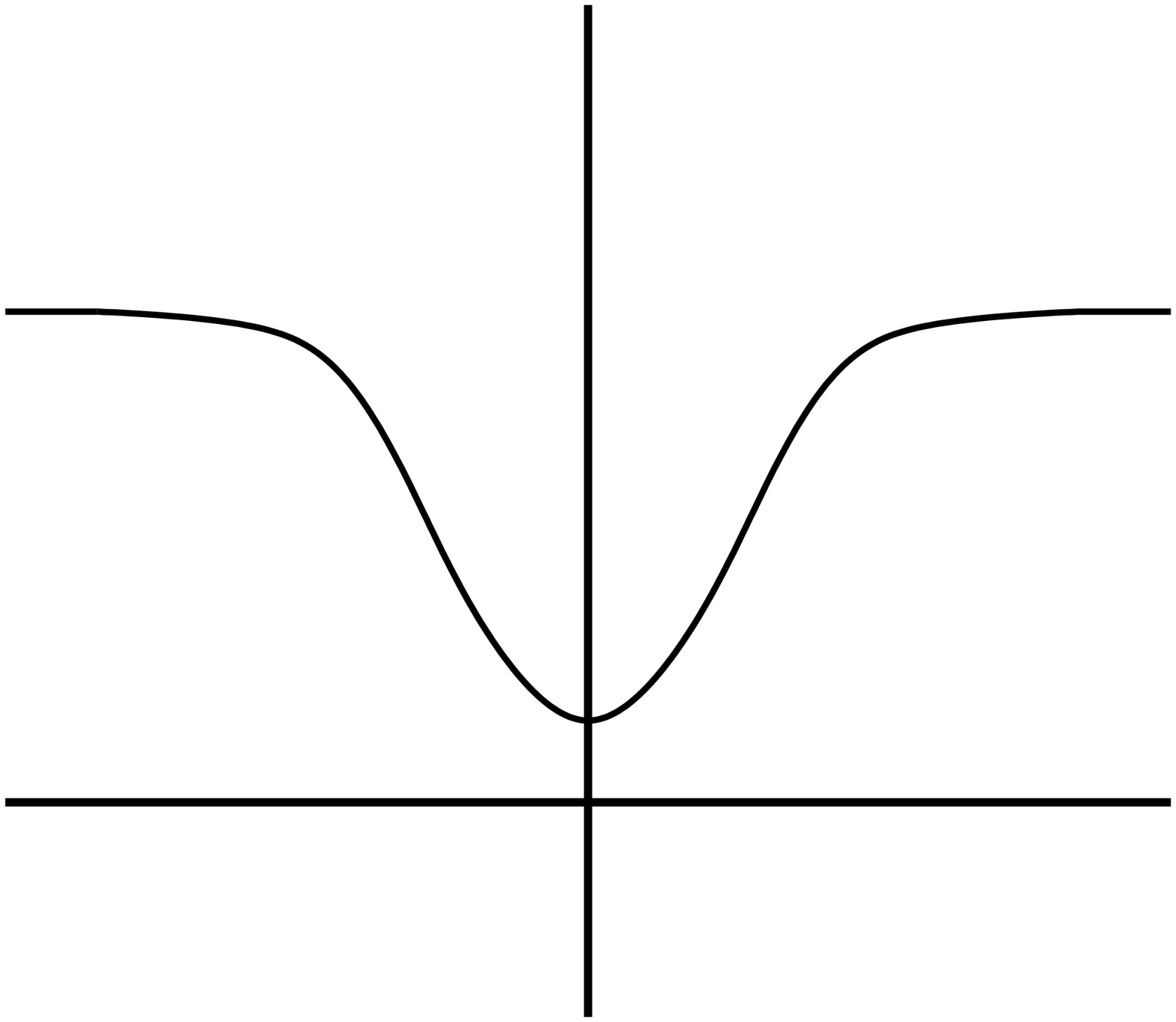}
\end{center}
    \caption{Acceptable potentials within effective
    descriptions. Left hand side: The physical potential from standard
    non-linear sigma models as non-holomorphic part. Right hand side: Possible potential in our
    class of models.}
    \label{fig:potentials}
\end{figure}
When representing the non-holomorphic part of the action by a standard
non-linear sigma-model
(cf.\ eq.\ \eqref{eq:geomeffact}) a stable potential for both, the auxiliary as well as the
physical fields cannot be obtained. As by its construction the non-linear sigma model is the most
general Lagrangian obeying all symmetries, we have to weaken some conditions
compared to this approach. This concerns the understanding of a stable
potential. We insist on the potential being bounded from below and having an
unique absolute minimum, identified with the physical minimum (we do not
consider models with a quantum moduli space, as the classical moduli space
must getting lifted when exploring the hysteresis line \cite{bergamin01}). In contrast to
\eqref{eq:geomeffact} we however accept potentials that become flat above some
value of the fields (cf.\ figure \ref{fig:potentials}). This is motivated by
the following observation: Our description is valid below some energy-scale
$\Lambda$ (or equivalently within restricted local excitations of the sources)
as well as within a certain range of the global sources, only. A breakdown of
the description outside of this range is rather a necessity than just a
possibility. This breakdown can either be seen in the momentum-expansion or in
the potential. However we have to insist on a potential bounded from below, as
the physical minimum is defined as the absolute minimum of the effective
potential (for a detailed discussion of this point see
\cite{bergamin01}). Thus a potential becoming flat above some scale of the
fields is indeed the most general situation. Of course this scale has now
direct physical implications. As we discuss in
this paper the general model without any reference to a concrete application,
we do not attend to this point within this
work.

This general choice of acceptable potentials as well as other steps of our
construction will lead to an ansatz for the effective action, which is not an
acceptable (classical) field theory for its own. It is of particular
importance in the discussion on hand that this need not be the case -- in fact
we will see that any acceptable description of the effective action must
disobey important features of classical supersymmetric field theories. From the point of view of the underlying
quantum field theory these
non-supersymmetric aspects of the effective description will turn out to be in
perfect agreement with all symmetries. We emphasize that supersymmetry (or any other symmetry
realized in the system) can be understood from this point of view, only. Many
problems in the description of dynamically broken supersymmetry and its
hysteresis line can be resolved by dropping the unfounded assumption that such
a model must be described by the classical supersymmetric non-linear sigma
model of equation \eqref{eq:geomeffact}. In this context it is important to
note that our model should be understood
in a complete non-perturbative study of quantum field theories,
only. Clearly a perturbative analysis of the same models must be compatible
with standard superspace geometry even when formulated in terms of the same
operators as used in the non-perturbative region.

To avoid misunderstandings we should shortly comment on the notion of an
effective action used in this work.
\begin{itemize}
\item  We consider as effective action a quantum effective action obtained by
  Legendre transformation. This action is a purely classical object, where all
  quantum effects have been summed up. We emphasize that we cannot escape the
  difficulties discussed so far by switching to an
  alternative low-energy description, especially to a Wilsonian low-energy
  effective action. Indeed, some of the mentioned problems may be absent in
  the Wilsonian action, but instead of solving them it simply gets rid of an
  important part of the dynamics by introducing an arbitrary infrared
  regulator. In consequence the quantities appearing in the Wilsonian action (e.g.\
  the coupling constant or the low-energy fields) are not physical, but the
  physical quantities are found after a (perturbative \emph{and} non-perturbative)
  renormalization step, only. It has been discussed in detail in \cite{bergamin01} why the
  Wilsonian action alone cannot serve as an alternative to the quantum
  effective action.
\item Dynamical auxiliary fields appear in a quite different context as well: If the non-linear
model is not seen as a purely classical object (as in this paper) but still
has quantum degrees of freedom the auxiliary fields become dynamical without
introducing any kinetic terms by hand. This is the direct consequence of
supersymmetry Ward identities, which read for a chiral superfield
\begin{equation}
  \lomega T \bar{F}(x) F(x') \romega = \Box \lomega T \bar{\varphi}(x)
  \varphi(x') \romega\ .
\end{equation}
In a linear theory this relation defines $F$ to be an auxiliary field, in a
non-linear theory it makes $F$ dynamical. Though this is a quite different
mechanism than the one proposed in this paper, our discussion is not
irrelevant for this case. The important observation is the following: If $F$
is a dynamical field, the potential must be bounded from below without
elimination of the auxiliary fields. Whether the origin of the kinetic terms
in $F$ are quantum dynamics or just some terms written into the Lagrangian by
hand is completely irrelevant. From this point of view mainly the discussion
of the new interpretation of the superpotential given in this paper is valid
for a non-linear quantum model as well. As an example, the spectrum of the
superpotential by Veneziano and Yankielowicz \cite{veneziano82} is getting
changed compared to that work in our models as well as in a quantum treatment
of this Lagrangian. In both cases the elimination of the $F$ field
performed in \cite{veneziano82} is forbidden and e.g.\ the second derivative
of the superpotential is no longer proportional to the mass of $\varphi$
(nevertheless it does still define a mass term of $\psi$). Although our
discussion takes place on a completely classical level this aspect of the
problem is of fundamental importance in a non-linear quantum model as well.
\end{itemize}
\subsection{Constraint K\"{a}hler Geometry with Dynamical Auxiliary Fields}
A physical potential of the \second\ generation auxiliary field within the
ansatz \eqref{eq:geomeffact} would be possible with $\metr < 0$ (equivalent to
$c_0 < 0$ in \eqref{eq:unstable}), only. It is easy to check, that the
instabilities caused by this ``wrong'' sign cannot be removed. On the other
hand, a physical potential is possible, if and only if the highest power in
$F$ comes together with a positive sign in the effective potential. The only
way out is thus an effective Lagrangian containing higher powers in $F$ (at
least $|\bar{F}F|^2$). Therefore we have to define a new superfield, where $F$
appears as \emph{lowest} component. Starting from the effective superfield
$\Phi$ we construct further dependent effective fields
according to\footnote{If the mass dimension of $\Phi$ is not 1 the modified
  sequence $\Psi_0 = \Phi^{1/d}$, $\Psi_n = \bar{D}^2 \bar{\Psi}_{n-1}$ should be
  considered. An example is $N=1$ SYM discussed in ref.\
  \cite{Bergamin:2003ub}.}
\begin{align}
\label{eq:sequdef}
  \Phi_0 &\equiv \Phi\ , & \Phi_n &= \bar{D}^2 \bar{\Phi}_{n-1}\ , & \Phi_{2n} &=
  (-1)^n \Box^n \Phi_0\ , & \Phi_{2n+1} &=
  (-1)^n \Box^n \Phi_1\ .
\end{align}
The most general effective Lagrangian of the chiral field $\Phi$ is now given by\footnote{It had been realized in \cite{Shore:1983kh} that the
  symmetries of $N=1$ SYM allow terms similar to the actions \eqref{eq:intro1}
  or \eqref{eq:nonlocalintro}. However, a detailed analysis of the system had
  not been given therein, which led the author to conclusions different from
  the ones presented in this work.}
\begin{equation}
  \label{eq:intro1}
  \mathcal{L} = \intd{\diff{^4x}} \biggl( \intd{\diff{^4 \theta}} A(\Phi_0,\Phi_1 , \bar{\Phi}_0, \bar{\Phi}_1) + \bigl(
  \intd{\diff{^2 \theta}} H(\Phi_0) + \hc \bigr) \biggr) \ .
\end{equation}
As this effective Lagrangian allows the construction of a potential bounded
from below in \emph{all} fields (as motivated in the previous section), we can
--in contrast to \eqref{eq:geomeffact}-- relax the constraint of $A$ and $H$
being polynomial functions in $\Phi_0$ and $\Phi_1$. Instead they are allowed to
 include explicit space-time derivatives. To make contact to the standard
 formulation of effective Lagrangians in terms of a K\"{a}hler- and a
 superpotential we may define a related action as
\begin{equation}
  \label{eq:nonlocalintro}
    \mathcal{L} = \intd{\diff{^4x}} \biggl( \intd{\diff{^4 \theta}} K(\Phi_n,
  \bar{\Phi}_n) - \bigl(\intd{\diff{^2 \theta}} W(\Phi_0) + \hc \bigr) \biggr)\ .
\end{equation}
Here the index $n$ runs from zero to infinity according to equation
\eqref{eq:sequdef}. $K$ and $W$ are polynomial functions, i.e.\ before using
the constraint \eqref{eq:sequdef} they describe the standard K\"{a}hler- and
superpotential. Both actions \eqref{eq:intro1} and \eqref{eq:nonlocalintro}
have the same effective potential. However, \eqref{eq:nonlocalintro} is a
restricted version of \eqref{eq:intro1} as it does not include all derivative
terms of the latter. To obey fundamental stability conditions certain constraints among the
different geometrical objects in \eqref{eq:intro1} and
\eqref{eq:nonlocalintro} have to be fulfilled.

Before going into a detailed discussion of
\eqref{eq:intro1}/\eqref{eq:nonlocalintro} the two important characteristics
of these actions are set out again:
\begin{enumerate}
\item The \second\ generation ``auxiliary'' field $F$, being the highest
  component of $\Phi = \Phi_0$, appears as \emph{lowest} component of
  $\Phi_1$. Thus both scalar fields, $F$ and $\varphi$, can appear with any
  power in the effective potential.
\item The equation of motion of $F$ is not algebraic. Thus this field cannot
  be eliminated. Instead it must be treated as an independent physical degree
  of freedom. 
\end{enumerate}

To discuss the effective potential of
\eqref{eq:intro1}/\eqref{eq:nonlocalintro} we can suppress all $\Phi_n$
($n>2$), or equivalently, restrict the functions $A$ and $H$ to be polynomials
in the fields. The corresponding action is given by
\begin{equation}
  \label{eq:geomeffactII}
  \mathcal{L} = \intd{\diff{^4x}} \biggl( \intd{\diff{^4 \theta}} K(\Phi_0, \Phi_1;
  \bar{\Phi}_0, \bar{\Phi}_1) - \bigl(\intd{\diff{^2 \theta}} W(\Phi_0) + \hc
  \bigr) \biggr)\ ,
\end{equation}
and  in a first step we analyze its effective potential, the dynamics around
the minimum of $V$ are discussed in a second
step. Integrating out superspace the potential becomes (all relevant
superspace integrals used in the following are listed in the appendix):
\begin{equation}
  \label{eq:effpot1}
  \begin{split}
    V &= - \metr F \Bar{F} + \half{1}
    \metr[, \Bar{\varphi}] F (\Bar{\psi} \Bar{\psi}) + \half{1}
    \metr[, \varphi] \Bar{F} (\psi \psi) -
    \inv{4} \metr[, \varphi \Bar{\varphi}] (\psi
    \psi) (\Bar{\psi} \Bar{\psi}) \medsp
    &\quad  + F W_{, \varphi}
    + \Bar{F} \bar{W}_{, \Bar{\varphi}} - \half{1}  (\psi
    \psi) W_{, \varphi \varphi} - \half{1}  (\Bar{\psi}
    \Bar{\psi}) \bar{W}_{, \Bar{\varphi} \Bar{\varphi}}
  \end{split}
\end{equation}
This potential is \emph{not} equivalent to the one of the model
\eqref{eq:geomeffact}, as the K\"{a}hler metric $\metr$ is a function of both
scalar fields, $F$ and $\varphi$
\begin{equation}
  \label{eq:metrdep}
  \metr = \metr(\varphi, F; \bar{\varphi}, \bar{F})\ .
\end{equation}
This dependence can include arbitrary powers in $\varphi$ as well as in
$F$. Analyzing the potential we insist on minima in \emph{all} three fields
$\varphi$, $\psi$ and $F$ as condition of the ground-state, a maximum for $F$
is not acceptable due to the dynamics of this field. The minima for $F$ are
found at
\begin{equation}
\label{eq:Fmincond}
\begin{split}
  \delta_F V\bigl. (\varphi, F; \bar{\varphi}, \bar{F})\bigr|_{F = F_0} &= -
  \bigl. \metr \bar{F} -  \metr[, F] \bar{F} F + \bwphi \bigr|_{F = F_0} = 0\ ,
  \medsp
  \delta_F \delta_{\bar{F}} V\bigl. (\varphi, F; \bar{\varphi},
  \bar{F})\bigr|_{F = F_0} &= \bigl. - \metr - \metr[, F \bar{F}] \bar{F} F -
  (\metr[, F] F + \metr[, \bar{F}] \bar{F} ) \bigr|_{F = F_0} > 0\ .
\end{split}
\end{equation}
If supersymmetry is unbroken the solution $F = 0$ must be the minimum of the
potential and thus the minimum condition $\delta_F \delta_{\bar{F}} V |_{F = F_0} > 0$
becomes $\metr < 0$. This leads to an unstable kinetic term for $\varphi$. We
could try to fix this by a mixing of $\varphi$ with $\bar{F}$ through
$g_{\varphi F} \neq 0$. From the point of view of fundamental stability
properties there may exist acceptable models of this type, but they are
certainly irrelevant in any physical application: The above equations tell us
that the potential for $\varphi$ must be completely flat for $F_0 = 0$. Thus
there exists neither the possibility of a vacuum expectation value for this
field (chiral symmetry breaking) nor for a mass term. Thus a model of this
type could never generate a mass gap.

 In the light of our discussion of the
last section this means that unbroken supersymmetry does not allow to deform
the flat potential of $\varphi$ as sketched in figure
\ref{fig:potentials}. Together with broken supersymmetry this deformation is
possible, but obviously any potential of this type becomes flat in the field
$\varphi$ for a (non-minimum) solution $F = 0$. At this point we use of the
``semi-stable'' potentials.

A relevant simplification in the discussion of broken supersymmetry is the
constraint on $F_0$, the latter must be real and positive, as it is the order
parameter of supersymmetry breaking. The Goldstino
is represented by $\psi$, as this is the field transforming into the order
parameter under supersymmetry transformations. For simplicity we set the
superpotential to zero and the minimum of the potential in F is found to be at
\begin{equation}
  F_0 = - \frac{\metr}{\metr[, F]}\ .
\end{equation}
The metric $\metr$ itself is real and thus the same applies to $\metr[, F]$
and we find that either $\metr$ or $\metr[, F]$ are smaller than zero\footnote{of
course all
quantities have to be evaluated in the minimum $F = F_0$, $\varphi = \varphi_0$,
$\psi \equiv 0$, but we suppress the index 0 wherever the meaning is
obvious}. If the K\"{a}hler potential depends on the real combination $\bar{F} F$ only, the vacuum expectation value reduces to
\begin{equation}
  (\bar{F} F)_0 = - \frac{\metr}{\metr[,F \bar{F}]}\ .
\end{equation}
A positive mass term for $\bar{F} F$ moreover tells us that
\begin{equation}
  \metr - \metr[, F \bar{F}] (\frac{\metr}{\metr[, F]})^2 > 0\ .
\end{equation}
Analogously we find the conditional equation for the vacuum expectation value
of $\varphi$ and the corresponding mass term
\begin{align}
  \bigl.\metr[, \varphi]\bigr|_{\varphi = \varphi_0} &= 0\ , & \bar{F} F
  \metr[, \varphi \bar{\varphi}] &= \bigl. \metr[, \varphi \bar{\varphi}]
  (\frac{\metr}{\metr[, F]})^2 \bigr|_{\varphi = \varphi_0} < 0
\end{align}
and thus $\metr[, \varphi \bar{\varphi}] < 0$. This last constraint follows
from the stability of the $\psi$ potential as well. In addition we see that
$\psi$ is indeed a massless Goldstone particle as 
\begin{equation}
\label{eq:psimass}
  m_\psi \propto \metr[, \bar{\varphi}] \bar{F}_0 = 0\ .
\end{equation}

More involved than the straightforward discussion of the potential is the
derivation of consistent dynamics around the minimum described
above. Evaluating the momentum expansion \eqref{eq:pexp1}-\eqref{eq:pexp4} at the minimum we
find the following bilinear terms
  \begin{align}
    \mathcal{L}^{(1)} &= - \half{i} \metr \psi \sigma^\mu \lrpartial_\mu
    \bar{\psi}\ ,\\[2ex]
    \begin{split}
      \Ltext{sc}^{(2)} &= - \metr \partial_\mu \bar{\varphi} \partial^\mu
      \varphi + g_{F \bar{F}} \partial_\mu \bar{F} \partial^\mu F  - \bigl( \metr
      \frac{g_{\varphi F, \bar{F}}}{\metr[,\bar{F}]}
      \partial_\mu \bar{F} \partial^\mu \varphi + \hc \bigr)  \spsp
      &\quad + \bigl( ( 2 g_{\varphi F} - \metr \frac{g_{\varphi F, F}}{\metr[,F]} )
      \partial_\mu F \partial^\mu \varphi + \hc \bigr) - \bigl( \metr
      \frac{g_{\varphi \varphi, F}}{\metr[, F]} \partial_\mu \varphi
      \partial^\mu \varphi + \hc \bigr)\ ,
    \end{split}\\[2ex]
    \Ltext{fer}^{(2)} &= ( g_{\varphi F} - \half{1} \metr \frac{g_{\varphi F,
    F}}{\metr[,F]} ) \psi \Box \psi + ( g_{\bar{\varphi} \bar{F}} -
      \half{1} \metr \frac{g_{\bar{\varphi} \bar{F},
    \bar{F}}}{\metr[,\bar{F}]} ) \bar{\psi} \Box \bar{\psi}\ ,\\[2ex]
    \mathcal{L}^{(3)} &= - \half{i} g_{F \bar{F}} \psi \sigma^\mu
    \lrpartial_\mu \Box \bar{\psi}\ ,\\[2ex]
    \mathcal{L}^{(4)} &= g_{F \bar{F}} \Box \bar{\varphi} \Box \varphi\ .
  \end{align}
If $\metr > 0$ $\Ltext{sc}^{(2)}$ shows potential instabilities in the kinetic
term of $\varphi$. The existence of positive eigenvalues depends on the
details of $g_{\varphi F}$, nevertheless we can distinguish the following two
alternatives:
\begin{description}
\item[$\mathbf{\metr > 0}$] In this case we find the minimum of the potential
  to be negative: $V_0 < 0$. In the classical and perturbative region such a
  potential would not be compatible with supersymmetry, but it has been
  pointed out in \cite{bergamin01} that this may happen in the
  non-perturbative region. The application of this type of models could be
  important in a theory with non-vanishing Witten index, if the corresponding
  state shall be part of the Hilbert space.
\item[$\mathbf{\metr < 0}$] This is the somehow more natural solution, as the
  positivity of $F \approx \lomega {T^\mu}_\mu \romega$ comes together with $V_0
  > 0$.
\end{description}
In the following we demonstrate that stable dynamics can be introduced among
both types of minima at least for some finite range of $p^2$. The two types
have relevantly different characteristics and are strictly separated from each
other, as $\metr = 0$ is not a consistent solution. To arrive at stable
dynamics we have to add different terms including explicit space-time derivatives to the
Lagrangian. Its most general form would thus be the action
\eqref{eq:intro1}/\eqref{eq:nonlocalintro}. But these additional terms do not
contribute to the effective potential and thus play a special role. The main
reason to consider a quantum effective action is to
find the minimum of the effective potential. For this task it is sufficient to
show that for a certain model stable dynamics can be introduced, its most
general form is not of main interest. The following simpler considerations
are thus sufficient in the present context.
\subsubsection{Dynamics with $\mathbf{\metr > 0}$}
To ensure stable $p^2$ fluctuations we add $\mathcal{L}_{c}$
of \eqref{eq:cadd} to the Lagrangian. To get the correct sign for $\varphi \Box \bar{\varphi}$
we impose
\begin{equation}
  c_1 > \frac{\metr}{(\bar{F} F)_0} = \frac{(\metr[, F])^2}{\metr}\ .
\end{equation}
If $g_{F \bar{F}}$ has the wrong sign, $e$ in \eqref{eq:eadd} should be
chosen appropriately. Considering the off-diagonal terms, all of them can be
canceled by choosing $d$, $f$ and $g$ correctly in
\eqref{eq:dadd}-\eqref{eq:gadd}, as this allows independent coefficients for
all combinations appearing in $\mathcal{L}^{(2)}$.

Among the even powers in the momenta
  we see that $\mathcal{L}^{(4)}$ is now unstable. We can correct this by
  choosing $c_2 > 0$ with
  \begin{equation}
    c_2 > \inv{(\bar{F} F)_0} \bigl( g_{F \bar{F}} + c_1
     (\bar{\varphi} \varphi)_0 \bigr)\ .
  \end{equation}
Again additional contributions may cancel off-diagonal terms. At this point the term
$p^6$ is unstable, which could be changed by an appropriate choice of
$c_3$. This way we arrive at a Lagrangian with stable $p$-expansion up to any
given order. Denoting by $\Lambda$ the typical scale of the theory (scale of
supersymmetry breaking) and assuming $\metr = \mathcal{O}(1)$ we get the following list
\begin{align*}
  \varphi_0 &= \mathcal{O}(\Lambda) & F_0 &= \mathcal{O}(\Lambda^2) & c_k &=  \mathcal{O}(\Lambda^{-(k+2)})
\end{align*}
Although we need a cancellation of the terms in $c_i$ by terms in
$c_{i+1}$ it is not surprising that higher order terms are
indeed suppressed for $p^2 \ll \Lambda^2$ as all tunings are of order one.

Two important points of the above construction should be clarified here:
\begin{itemize}
\item The reference point of any scaling argument is now the minimum found in
  eqs.\ \eqref{eq:Fmincond}-\eqref{eq:psimass}. Thus dominance or suppression
  of any term cannot be compared with the situation of a static auxiliary
  field: Indeed, the vacuum of a situation with static auxiliary field ($F_0 =
  0$) is at a distance of $\mathcal{O}(\Lambda)$ from the correct minimum. At
  this point, the momentum-expansion discussed here breaks down. This again
  illustrates that models with dynamical \second\ generation ``auxiliary''
  fields must be discussed without any reference to models with static
  \second\ generation auxiliary fields.
\item Most terms including explicit space-time derivatives added above exist
  in the formulation \eqref{eq:intro1}, only. The instabilities may be lifted
  in the formulation \eqref{eq:nonlocalintro} using more complicated
  structures. But apart from its nice mathematical construction there exists
  no reason to prefer \eqref{eq:nonlocalintro} compared to
  \eqref{eq:intro1}. Thus we do not go into the details of this problem.
\end{itemize}

Before we go over to the discussion of the case $\metr < 0$ we want to make
some comments. As mentioned in the previous section our Lagrangian
can be seen as an effective description of some more
  fundamental theory, only. This can be illustrated by means of several
  properties of the model:
  \begin{itemize}
  \item It is well known that  the order parameter of 
    spontaneous supersymmetry breaking is directly related to the coupling of the goldstino
    and is restricted to positive values \cite{salam74}. We could try to construct the
    supercurrent of our model and read off the above quantities. At first
    sight these relations seem to be broken by the models discussed here: The
    typical representative of the order parameter is the
    vacuum expectation value of the energy-momentum tensor. From
    $(\mathcal{L})_0 = \metr (F \bar{F})_0$ we find $(T_{\mu \nu})_0 = -
    g_{\mu \nu} \metr (F \bar{F})_0 < 0$, but obviously this quantity has
    nothing to do with a goldstino coupling.

    To understand this behavior one should notice that the positivity
    of supersymmetric potentials is actually a constraint on the
    maximum of the auxiliary field potential, the latter must be positive
    semi-definite. By eliminating the auxiliary field the system is put on top
    of the auxiliary field potential, which becomes the minimum of the
    physical potential. In our model we minimize the potential in all fields
    and the positivity property is lost. 

    Nevertheless, the correct goldstino coupling is still realized: From the
    point of view of an effective theory it is determined
    from the order parameter of the underlying quantum field theory, which will
    typically be equivalent to the \second\ generation auxiliary field
    $F$. The goldstino coupling must be in agreement with the restrictions
    from the current algebra of this underlying quantum field theory, only. We cannot
    expect that similar relations from the effective theory have a direct
    (physical or mathematical) interpretation.
    \item In contrast to standard
    supersymmetry-breaking models we did not break supersymmetry by a splitting of the masses
    of the physical fermion and boson states, $\psi$ and $\varphi$. In a
    physical application $\varphi$ will typically be a massive state, but one
    should notice that we can break supersymmetry even with massless $\varphi$
    (an example is given in section \ref{sec:examples}). Instead we have arrived at a non-zero vacuum
    expectation value of the auxiliary field just by
    manipulating the potential thereof. This still generates the typical
    transformation-rule of a goldstino for $\psi$: $\delta_\alpha \psi_\beta =
    i \epsilon_{\alpha \beta} F_0 + \mbox{local}$ and the effective theory
    is in agreement with all current algebra relations of the fundamental theory.
  \item Closely related to the above observations is the atypical form of the
    potential with $\itindex{V}{eff}(\Phi_0) < 0$. It has been discussed in
    \cite{bergamin01} that such a potential need not contradict supersymmetry,
    but its realization needs the presence of non-perturbative
    non-semiclassical effects. Our model leads to an effective description of
    this type of supersymmetry breaking by means of a Lagrangian written in
    superspace. Thus by its definition the current-algebra
    relations of this model (taken for its own) cannot be consistent with
    standard results from classical supersymmetric field theories.
  \item The price we paid to arrive at the model is a wrong sign in the
    $p$-fluctuations of the goldstino. It is easy to check that all terms of odd
    order in the momentum have the same ``wrong'' sign. This may look
    unesthetic but one should
    rate any effective model according to stability conditions and correct realization
    of symmetries, only. From this point of view the wrong sign of the kinetic term is
    acceptable, it does not introduce instabilities but can be removed by
    interchanging positive and negative frequencies of the
    goldstino. Nevertheless this sign could be problematic if we try to
    couple the model to additional matter fields.
  \item Without reference to an underlying theory, the model violates the
    equality of bosonic and fermionic degrees of freedom. Again the underlying
    theory may resolve this: The equality
    obviously holds on the level of the (quantum-)field content of the
    classical fields in $\Phi$. Whether the equality is realized on the level
    of $\Phi$ by the suggestive solution with $\varphi$, $\psi$ physical and
    $F$ auxiliary or not is not at all obvious, physical $\varphi$ and $F$
    need not stand in contradiction to the equality as they are usually
    subject to constraints from the fundamental theory.

    To prevent
    misunderstandings we notice again that the treatment of the \second\
    generation ``auxiliary'' fields is completely independent from the \first\
    generation, i.e.\ it does not change when using Wess-Zumino gauge on the
    level of the fundamental theory. This may on the contrary motivate the
    above statement: By eliminating the fundamental auxiliary field, all
    fields of the effective superfield are built up from solely physical
    fields. Inspecting e.g.\ the effective superfield of SYM, the constraint
    on its highest component being auxiliary looks completely arbitrarily. Our
    construction shows how to avoid this at least for a certain class of models.
  \end{itemize}
\subsubsection{Dynamics with $\mathbf{\metr < 0}$}
If $g_{F \bar{F}} > 0$ the
fluctuations to order $p^2$ can be stable even when choosing $c_1 = 0$. As
before all off-diagonal terms can be canceled in $\mathcal{L}^{(2)}$. Again $\mathcal{L}^{(4)}$
is unstable and we choose appropriate values for $c_2$ or similar higher
derivative terms. But in contrast
to the above discussion it is impossible to write down a Lagrangian with a
consistent (though non-convergent) $p$-expansion for any value of the
momentum. While $\mathcal{L}^{(1)}$ has now the standard sign, this does not
apply to the other terms of odd order in the momentum. These wrong signs are
closely related to fundamental characteristics of the geometry of chiral
fields and cannot be changed with the techniques presented in this paper. Thus
the expansion breaks down at some $p = p_0$ of the order of
$\Lambda$. As a peculiarity this is not a tachionic instability (we are
still able to remove all of them), but the goldstino becomes static at this
point and interchanges positive and negative frequencies for momenta larger
than $p_0$.
\subsubsection{Including the Superpotential}
Finally models with a superpotential are shortly discussed. Using the potential
\eqref{eq:effpot1} its minima are found at
\begin{equation}
  \label{eq:Fominima}
  F_0 = - \inv{\metr[, F]} \bigl( \metr \pm \sqrt{(\metr)^2 + 4 \metr[, F]
  \wphi} \bigr)\ .
\end{equation}
Independent of the sign of $\metr$ the $+$-solution is the absolute minimum of
the potential. For $\metr > 0$ one finds the conditions
\begin{align}
  \metr[, F] &< 0\ , & \wphi &< - \frac{(\metr)^2}{4 \metr[, F]}\ ,
\end{align}
if $\metr[, F]$ and $\wphi$ are both real in the minimum (there exist of
course solutions where these quantities are complex, but $F_0$ still
real). For $\metr < 0$ the sign of $\metr[, F]$ is not fixed as long as
$\wphi$ is negative and
\begin{equation}
  |\wphi| < \frac{(\metr)^2}{4 |\metr[, F]|}\ .
\end{equation}

The generalization of the $\psi$ and $\varphi$ potentials is
straightforward. The goldstino is again massless and stable if $\metr[,
\varphi \bar{\varphi}] < 0$. The vacuum expectation value of $\varphi$ is
determined by
\begin{equation}
  \bigl. - \metr[, \varphi] (\bar{F} F)_0 + \wphi[\varphi] F \bigr|_{\varphi =
  \varphi_0} = 0\ .
\end{equation}
In the $p$-expansion some constraints look more complicated when introducing a
superpotential, but nothing changes fundamentally. We will illustrate its
effect at hand of an example, a more realistic application is discussed in \cite{Bergamin:2003ub}.


\section{Examples}
\label{sec:examples}
As illustration we provide some examples of the model developed in the
previous section. They should
give some feeling about the possibilities within this approach and do not have
a specific application in a real model. For simplicity we introduce the
notation $\Phi_0 = \Phi$, $\Phi_1 = \Psi$.

The simplest consistent constraint K\"{a}hler potential is given by
\begin{equation}
  K(\Phi, \Psi; \bar{\Phi}, \bar{\Psi}) = \bar{\Psi} \Psi + \bar{\Phi} \Phi
  I(\Psi, \bar{\Psi})\ ,
\end{equation}
its minimum found at
\begin{align}
  F_0 &= - \frac{I(F, \bar{F})}{\partial_F I(F, \bar{F})}\ , & \varphi_0 &= 0\ .
\end{align}
As $\varphi_0 = 0$, the kinetic terms of the scalars are automatically diagonal
to all orders. Moreover the ones of the goldstino appear with odd powers of the momentum,
only. For $\metr > 0$ we can choose e.g.\ $I = g_2 + g_4 \bar{F} F$ with $g_2
> 0$ and $g_4 < 0$, the standard Mexican-hat potential. The minimum of the
potential is given by $(\bar{F} F)_0 =
\frac{g_2}{2 |g_4|}$ and $(\metr)_0 = \half{g_2}$ and we derive the
stability constraints
\begin{align}
  c_1 &> |g_4| & c_2 &>  2 \frac{|g_4|}{g_2}\ .
\end{align}
With this choice the Lagrangian has a stable $p$-expansion stopping at order
$p^4$. Higher order derivatives can be introduced by choosing $c_2$, $c_3$
\ldots and the related series from $\eqref{eq:eadd}$ non-zero. A similar
phenomenology has a model with $\metr < 0$, e.g. $ I = -
(\frac{\alpha}{(\bar{F} F)^2} + \beta)$, $\alpha > 0$ and $\beta > 0$, with
\begin{align}
\label{eq:maslesstwo}
  V &= \frac{\alpha}{\bar{F} F} + \beta \bar{F} F\ , & (\bar{F} F)_0 &=
  \sqrt{\frac{\alpha}{\beta}}\ , & \metr&= - 2 \beta\ .
\end{align}
The momentum expansion again stops at order $p^4$ and we have to choose $c_2
> \sqrt{\frac{\beta}{\alpha}}$. Notice the difference in the signs of
$\mathcal{L}^{(1)}$ and of $\mathcal{L}^{(3)}$. These two examples illustrate
the fact that we can break supersymmetry without a split of the masses of
$\varphi$ and $\psi$, but they are both massless.

We can immediately generalize the above example to models of the type
\begin{equation}
  K(\Phi, \Psi; \bar{\Phi}, \bar{\Psi}) = f_2 \bar{\Psi} \Psi + \bar{\Phi} \Phi
  I(\Psi, \bar{\Psi}) + J(\Phi, \bar{\Phi})\ .
\end{equation}
These models have a minimal coupling between $\Phi$ and $\Psi$
with the potential
\begin{equation}
  V = - \bar{F} F \bigl( I(F, \bar{F}) + \partial_\varphi
  \partial_{\bar{\varphi}} J(\varphi, \bar{\varphi}) \bigr)\ .
\end{equation}
If in addition the vacuum expectation value of $\varphi$ vanishes, all
constraints on the dynamics are equivalent to the free model above. For
$\varphi_0 \neq 0$ however, the dynamics become more complicated. Even for the
simple minimal coupling of $\Psi$ to $\Phi$ we get off-diagonal terms in the
kinetic Lagrangian. An example of this type with $\metr > 0$ is the
double-Mexican-hat with the potentials
\begin{align}
  I(F, \bar{F}) &= g_2 + g_4 \bar{F} F\ , & J(\varphi, \bar{\varphi}) &=
  \frac{h_2}{4} (\bar{\varphi} \varphi)^2 + \frac{h_4}{9} (\bar{\varphi}
  \varphi)^3\ .
\end{align}
The potential has the minimum at
\begin{align}
  (\bar{\varphi} \varphi)_0 &= \frac{h_2}{2 |h_4|}\ , & (\bar{F} F)_0 &= -
  \inv{2 g_4} (g_2 + \frac{h_2^2}{4 |h_4|} )\ .
\end{align}
The complete $\mathcal{L}^{(2)}$ of this example reads:
\begin{equation}
  \label{eq:ex1L2}
\begin{split}
  \mathcal{L}^{(2)} &= - \bigl( \half{g_2} + \inv{8} \frac{h_2^2}{|h_4|}
  \bigr) \partial_\mu \varphi \partial^\mu \bar{\varphi} + \bigl(f_2 + g_4
  \frac{h_2}{2 |h_4|} \bigr) \partial_\mu F \partial^\mu \bar{F} \medsp
  &\quad - \Bigl( \bigl( \half{g_2} + \inv{8} \frac{h_2^2}{|h_4|}
  \bigr) \bigl(\frac{\bar{\varphi}}{F}\bigr)_0 \partial_\mu \bar{\varphi}
  \partial^\mu \varphi + \hc \Bigl) + \Bigl(g_4 (\bar{\varphi}\bar{F})_0
  (2 \partial_\mu F \partial^\mu \varphi + \psi \Box \psi)  + \hc \Bigr) \medsp
\end{split}
\end{equation}
In this equation $F_0$ must be real and positive, while $\varphi_0$ has a free
phase. As $f_2$ does not contribute to the potential we can assume without
loss of generality that $g_{F\bar{F}} > 0$. The diagonal sector then has
positive eigenvalues if $c_1 > g_4$. This will lead to new contributions
$\propto \partial_\mu \varphi \partial^\mu \bar{F}$, all of them can be
canceled by an appropriate choice of $d$ in \eqref{eq:dadd}. Finally we
choose in \eqref{eq:gadd}
\begin{equation}
  g = \frac{|g_4|}{2 (\bar{F}F)_0}\ ,
\end{equation}
which cancels the last expression of \eqref{eq:ex1L2}. This way we arrive at a
completely diagonal and bosonic $\mathcal{L}^{(2)}$.

In a similar way the higher order derivatives can be arranged. Adding a
superpotential leads to cubic equations in $F_0$, whose explicit form is not
very illuminating. Notice however that the cancellation of off-diagonal terms
and the resulting structure of the kinetic term matrix does not depend on the
absence of a superpotential. Especially one can again use $f_2 > 0$ to get two
positive eigenvalues. An interesting but simple
superpotential is $W = \frac{w_3}{6} \Phi^3$. The minimum $\varphi_0$ is then
found at
\begin{equation}
  (\bar{\varphi} \varphi)_0 = \inv{2 |h_4|} \bigl( h_2 -
  \inv{F_0} \frac{\varphi_0}{\bar{\varphi}_0} w_3 \bigr) 
\end{equation}
As $h_2$, $h_4$ and $F_0$ are all real the phase of $w_3$ directly determines
the phase of $\varphi$:
\begin{equation}
  \arg w_3 = - \half{1} \arg \varphi_0
\end{equation}

Finally we want to give an example with singular potential at the origin. For
simplicity we choose again the same shape of the $F$ and $\varphi$ potential,
namely
\begin{align}
  I &= - (\frac{\alpha}{(\bar{F} F)^2} + \beta) & J &= - ( \tilde{\alpha} \ln
  \frac{\varphi}{\Lambda} \ln \frac{\bar{\varphi}}{\Lambda} +
  \frac{\tilde{\beta}}{4} (\bar{\varphi} \varphi)^2 )\ ,
\end{align}
where all coupling constants are positive and $\Lambda$ denotes the scale of
the theory. The minima are found at
\begin{align}
  (\bar{\varphi} \varphi)_0 &= \sqrt{\frac{\tilde{\alpha}}{\tilde{\beta}}} &
  (\bar{F} F)_0 &= \sqrt{\frac{\alpha}{\beta + 2 \sqrt{\tilde{\alpha}\tilde{\beta}}}}
\end{align}
As in the same model with vanishing $\varphi_0$ (eq.\ \eqref{eq:maslesstwo}), the potential is positive
definite and thus $\metr < 0$. Without loss of generality we can choose
$g_{\bar{F} F} > 0$ and thus the diagonal part of $\mathcal{L}^{(2)}$ is
stable even for $c_1 = 0$. The term $\propto \partial_\mu \varphi \partial^\mu
\varphi$ vanishes again while the coefficients of the
$\varphi$-$F$ and $\varphi$-$\bar{F}$ kinetic terms are found to be:
\begin{align}
  2 g_{\varphi F} - \metr \frac{g_{\varphi F, F}}{\metr[,F]} &= -2 \bigl(\beta
      + \sqrt{\tilde{\alpha} \tilde{\beta}} \bigr) \frac{\bar{\varphi}_0}{F_0}
      &  - \metr
      \frac{g_{\varphi F, \bar{F}}}{\metr[,\bar{F}]} &= - 4 \bigl(\beta
      + \sqrt{\tilde{\alpha} \tilde{\beta}} \bigr) \frac{\bar{\varphi}_0}{F_0}
\end{align}
These off-diagonal terms can be canceled by choosing
\begin{align}
  d &= 2 \frac{\bigl(\beta
      + \sqrt{\tilde{\alpha} \tilde{\beta}} \bigr)^3}{\alpha^2} & g &=
      \half{1} \frac{\bigl(\beta
      + \sqrt{\tilde{\alpha} \tilde{\beta}} \bigr)^3}{\alpha^2}
\end{align}
Again we have to choose the parameters of the higher order derivatives
according to the discussion in the previous section.

We have illustrated in this section at hand of a few simple examples, how an
effective description of a non-perturbatively broken SUSY theory could be
realized. Of course the examples have been too simple for any realistic
application and from this point of view we conclude this section with a few
remarks:
\begin{itemize}
\item In a theory with dynamical ``auxiliary'' fields, the superpotential looses
  most of its power. This has a simple reason: As long as we can eliminate the
  auxiliary fields the superpotential
  produces automatically super-stable potentials (i.e.\ they are automatically
  positive-semidefinite). In a theory with dynamical auxiliary fields, the
  contrary is true: The superpotential alone is \emph{necessarily} unstable,
  as its holomorphic character in $\varphi$ cannot be changed through the equations of motion of
  $F$. We emphasize again that this fact does not depend on the origin of the
  dynamics. Here they came from explicit new terms written into the
  (classical) Lagrangian, but the same conclusion applies if they are an
  effect of quantum dynamics

  Nevertheless in a concrete application the superpotential is
  important. Veneziano and Yankielowicz \cite{veneziano82} have shown how to
  realize the anomalies of SYM in the superpotential. Of course a similar term
  must be present in a generalized construction of this effective action based
  on the Lagrangian \eqref{eq:intro1}/\eqref{eq:nonlocalintro} \cite{Bergamin:2003ub}.
\item Closely related to the above observations is the question of the
  spectrum of our models. In the above examples the Lagrangian respects the
  two global $\uone$ symmetries from the complex fields $F$ and
  $\varphi$ (except for the example including a superpotential). Consequently one ($\varphi_0 = 0$) or two Goldstone bosons are
  present in addition to the goldstino. At least one of them (associated with
  $F$) must be absent: $F$ is the order parameter of supersymmetry breaking,
  its value must be  real and positive and thus the direction of the vacuum is fixed. If
  $\varphi_0 \neq 0$ is related to chiral symmetry breaking, the corresponding
  Goldstone boson is absent as well. Such a potential can be realized
  together with holomorphic terms from a superpotential, as demonstrated in one example.
\item Finally it is important to note that a concrete application of our ideas
  relies on a quantum effective action, defined via a source extension of the
  classical system. Such a source extension breaks intrinsic supersymmetry but
  should preserve the latter extrinsically \cite{bib:mamink, bergamin01}. By taking the
  limit of constant sources we thus arrive at a system with softly broken
  supersymmetry and the (pseudo-)goldstino receives a mass. Therefore it will be
  important to include a spurion field into the description. Some comments on
  this problem for $N=1$ SYM are given in \cite{Bergamin:2003ub}, for a recent
  discussion of related problems in perturbation theory we refer to \cite{kraus01, kraus01:3, kraus02:1,Kraus:2002nu}.
\end{itemize}

\section{Summary and Conclusions}
We have introduced in this paper a class of low-energy descriptions of
supersymmetric gauge theories and demonstrated its features at hand of some
simple examples. Our
description is suitable for models where supersymmetry is broken dynamically
by non-perturbative non-semiclassical effects. The specific problem in the
construction of the low-energy approach lies in the realization of
supersymmetry: In a sense the model must be both, supersymmetric and
non-supersymmetric. The supersymmetric and non-supersymmetric aspects are:
\begin{itemize}
\item Supersymmetry is realized linearly in the sense of transformation rules
  between the different low-energy degrees of freedom. This behavior is
  included in the assumption of a correct identification of the low-energy
  degrees of freedom. In practice it simply means that our ansatz must be
  expressible as integrals over superspace.
\item Taken for its own the effective model must disobey standard
  characteristics of (classical and perturbative) supersymmetric models. This
  can be motivated as follows: Our model shall describe by means of a
  classical Lagrangian the behavior of a theory, in which important
  characteristics of classical and perturbative supersymmetry have been
  changed by non-perturbative effects. This classical (effective) Lagrangian
  must thus in certain aspects be non-supersymmetric. This especially
  includes:
  \begin{enumerate}
  \item The usual splitting of the fields into physical ones (with an equal
    number of bosonic and fermionic degrees of freedom) and into auxiliary
    fields does not hold. Instead the auxiliary fields turn into physical
    fields.
  \item The minimum of the potential can be negative. In classical and
    perturbative supersymmetry this is excluded by the analogy of the minimum
    of the potential and the vacuum expectation value of the energy-momentum
    tensor due to the non-renormalization theorem. The latter must be positive semi-definite from current algebra
    relations. In the non-perturbative region this relation can be broken and
    thus there exists no constraint on the value of the potential in its
    minimum.
  \item Supersymmetry is broken dynamically, but the Goldstino coupling on the
    level of the effective model is not realized as in classical and
    perturbative supersymmetry.
We should emphasize that all these non-supersymmetric characteristics can
be seen as non-supersymmetric from the point of view of the effective model
(taken for its own and not referring to the underlying theory), only. From the
point of view of the underlying theory all seeming discrepancies have its
definite interpretation. Especially it has been worked out in
\cite{bergamin01} that the possibility of dynamical auxiliary fields, of
infinite orders of derivatives and of a classically non-supersymmetric shape
of the potential are of fundamental importance.
  \end{enumerate}
\end{itemize}
The ansatz used in this work is consistent together with dynamically broken
supersymmetry, only. Typically we assume in this type of theories the
existence of a mass-gap where solely the goldstino lives below the
latter. One then might ask what our formulation gains compared to a low-energy
description of the goldstino, which can be found within non-linear realizations
of supersymmetry \cite{ivanov78,samuel83}. We think that there exists an
important difference between the two situations: Indeed our specific model
has to assume dynamically broken supersymmetry, but the motivation stems from
a quantity, whose existence does not depend on this question, the
quantum effective action. Our analysis together with the discussion of
\cite{bergamin01} shows that a single ansatz for the description of the
quantum effective action is probably insufficient, instead we need two, one
for broken and one for unbroken supersymmetry. Stability and consistency
conditions evaluated in both cases then should show, whether supersymmetry actually
breaks or not. But this program does not allow to replace the description of
the broken case by the above mentioned non-linear effective
goldstino-Lagrangian: The field-content is determined by the source-extension
of the system and supersymmetry must be realized on this set of
fields. Moreover the description must hold even for softly broken
supersymmetry, which can easily be included in our approach. From this point of view we think that our
description is more fundamental: Although the specific realization had to
assume broken supersymmetry it can be used within a program that does
not presume this behavior. Once we have found therefrom that
supersymmetry is actually broken, the above goldstino-Lagrangian may be a
simpler and more convenient way to describe the theory for vanishing
sources. For a more detailed discussion of this relation between
fundamental models describing the quantum effective action and effective
approaches describing the dynamics once we have found the ground-state, we
refer the reader to \cite{bergamin01}.

Many questions are still open. A concrete application to a physical system has
not been given within this work. Its application to $N=1$ SYM is discussed in
\cite{Bergamin:2003ub}. In addition the model is restricted to one classical
chiral field (the goldstino field). It would be interesting to see, whether
one can couple additional fields to the system, either from higher
supersymmetry or as Goldstone bosons from the breakdown of bosonic
symmetries. Finally the existence of the non-perturbative
non-semiclassical effects has been derived in \cite{bergamin01} from physical
arguments, only. Classical field-configuration that could lead to such an
effect are unknown. In other words: It would be interesting to turn the
non-perturbative non-semiclassical effect into a non-perturbative
semiclassical one.


%
\section*{Acknowledgement}
The authors would like to thank Ch.\ Rupp for interesting discussions on the
topic. One of us (L.B.) is grateful to M.\ Bianchi, F.\ Lenz, J.M.\ Pawlowski
and E.\ Scheidegger for illuminating
comments. This work has been supported by the Swiss National Science
Foundation (SNF). 
\appendix
\section{Appendix}
\label{sec:appendix}
In this appendix we list the complete momentum expansion of the
constraint double-K\"{a}hler model and of some other superspace integrals used
in the paper. The fundamental K\"{a}hler potential is given by:
\begin{equation}
\begin{split}
  \intd{\diff{^4 \theta}} K(\Phi_1, \Phi_2; \bar{\Phi}_1, \bar{\Phi}_2) & =
  g_{i \unj} \bigl( \partial_\mu \bar{ \varphi}^j \partial^\mu \varphi^i +
  \half{i} \psi^i \sigma^\mu \lrcovar_\mu \bar{\psi}^j + F^i \bar{F}^j
  \bigl)\medsp
  & - \half{1} g_{i \unj} \Gamma^{\unj}_{\unk\, \unl} F^i \bar{\psi}^k \bar{\psi}^l
  - \half{1} g_{i \unj} \Gamma^i_{kl} \bar{F}^j \psi^k \psi^l
  + \inv{4} g_{i \unj, k \unl} \psi^i \psi^k \bar{\psi}^j \bar{\psi}^l
\end{split}
\end{equation}
$\Phi_2 = \bar{D}^2 \Phi_1$ and thus
\begin{align}
  \Phi_1 &= \varphi + \theta \psi + \theta^2 F & \Phi_2 &= \bar{F} - i \theta
  \sigma^\mu \partial_\mu \bar{\psi} - \theta^2 \Box \bar{\varphi}
\end{align}
Denoting the components of the K\"{a}hler potential by $1 = \varphi$,
$\underline{1} = \bar{\varphi}$, $2 = \bar{F}$ and $\underline{2} = F$ we get
the momentum expansion
\begin{align}
\label{eq:pexp0}
  \mathcal{L}^{(0)} &= - V = \metr F \bar{F} - \half{1} \metr[, \varphi] F
  \bar{\psi} \bar{\psi} - \half{1} \metr[, \bar{\varphi}] \bar{F} \psi\psi +
  \inv{4} \metr[, \varphi \bar{\varphi}] \psi\psi\bar{\psi} \bar{\psi} \\[2ex]
  \begin{split}
\label{eq:pexp1}
    \mathcal{L}^{(1)} &= i \bigl( \half{1} \metr + \metr[, \bar{F}] \bar{F} -
    \half{1} \metr[, \bar{F} \bar{\varphi}] \bar{\psi} \bar{\psi} \bigr) \psi
    \sigma^\mu \partial_\mu \bar{\psi}  - i \bigl( \half{1} \metr + \metr[, F] F -
    \half{1} \metr[, F \varphi] \psi \psi \bigr) \partial_\mu \psi
    \sigma^\mu \bar{\psi} \medsp
    &\quad + \half{i} \bigl(\metr[, \bar{\varphi}] \partial_\mu \bar{\varphi}
    + \metr[, F] \partial_\mu F - \metr[, \varphi] \partial_\mu \varphi
    - \metr[, \bar{F}] \partial_\mu \bar{F} \bigr) \psi \sigma^\mu \bar{\psi}
  \end{split}\\[2ex]
  \begin{split}
\label{eq:pexp2}
    \mathcal{L}^{(2)} &= \metr \partial_\mu \bar{\varphi} \partial^\mu \varphi
    + g_{F \bar{F}} \partial_\mu \bar{F} \partial^\mu F + g_{F \varphi}
    \partial_\mu F \partial^\mu \varphi + g_{\bar{F} \bar{\varphi}}
    \partial_\mu \bar{F} \partial^\mu \bar{\varphi} \medsp
    &\quad + \half{1} g_{\varphi F} \psi \sigma^\mu \bar{\sigma}^\nu
    \lrpartial_\mu \partial_\nu \psi - \half{1} g_{\bar{\varphi} \bar{F}}
    \partial_\nu \bar{\psi} \bar{\sigma}^\nu \sigma^\mu \lrpartial_\mu
    \bar{\psi} \medsp
    &\quad + \half{1} \bigl( \metr[, F] \partial_\mu \bar{\varphi} +
    g_{\varphi F, F} \partial_\mu F - g_{\varphi F, \varphi} \partial_\mu
    \varphi - g_{F \bar{F}, \varphi} \partial_\mu \bar{F} \bigr) \psi
    \sigma^\mu \bar{\sigma}^\nu \partial_\nu \psi \medsp
    &\quad - \half{1} \bigl( g_{\bar{\varphi} \bar{F}} \partial_\mu \bar{\varphi} +
    g_{\bar{\varphi} \bar{F}, F} \partial_\mu F - \metr[, \bar{F}] \partial_\mu
    \varphi - g_{\bar{\varphi} \bar{F}, \bar{F}} \partial_\mu \bar{F} \bigr)
    \partial_\nu \bar{\psi}
    \bar{\sigma}^\nu \sigma^\mu  \bar{\psi} \medsp
    &\quad - (g_{\varphi F} F - \half{1} g_{\bar{\varphi} \bar{F},
    \bar{\varphi}} \bar{\psi}\bar{\psi}) \Box \varphi - (g_{\bar{\varphi}
    \bar{F}} \bar{F} - \half{1} g_{\varphi F, \varphi} \psi\psi) \Box
    \bar{\varphi} \medsp
    &\quad - \half{1} ( g_{\varphi F, F} F - \half{1} g_{\varphi F, \varphi F}
    \psi\psi) \partial_\mu \psi \sigma^\mu \bar{\sigma}^\nu \partial_\nu \psi + \metr[, F \bar{F}] (\psi \sigma^\mu \partial_\mu \bar{\psi})
    (\partial_\nu \psi \sigma^\nu \bar{\psi}) \medsp
    &\quad -
    \half{1} ( g_{\bar{\varphi} \bar{F}, \bar{F}} \bar{F} - \half{1} g_{\bar{\varphi} \bar{F}, \bar{\varphi} \bar{F}}
    \bar{\psi}\bar{\psi}) \partial_\mu \bar{\psi} \bar{\sigma}^\mu \sigma^\nu
    \partial_\nu \bar{\psi}
  \end{split}\\[2ex]
  \begin{split}
\label{eq:pexp3}
    \mathcal{L}^{(3)} &= \half{i} g_{F \bar{F}} (\partial_\mu \psi \sigma^\mu
    \Box \bar{\psi} - \Box \psi \sigma^\mu \partial_\mu \bar{\psi})\medsp
    &\quad + \half{i}  \bigl( g_{F \bar{F}, \bar{\varphi}}
    \partial_\mu \bar{\varphi} + g_{F \bar{F}, F} \partial_\mu F - g_{F \bar{F}, \varphi}
    \partial_\mu \varphi + g_{F \bar{F}, \bar{F}} \partial_\mu\bar{F} \bigr) \partial_\nu \bar{\psi} \bar{\sigma}^\nu \sigma^\mu
    \bar{\sigma}^\rho \partial_\rho \psi
    \medsp
    &\quad + i \bigl( g_{F \bar{F}, \bar{\varphi}} \Box \bar{\varphi} +
    \half{1} g_{F \bar{F}, \bar{F} \bar{\varphi}} \partial_\mu \bar{\psi} \bar{\sigma}^\mu \sigma^\nu
    \partial_\nu \bar{\psi} \bigr) \partial_\rho \psi \sigma^\rho \bar{\psi}
    \medsp
    &\quad - i \bigl( g_{F \bar{F},\varphi} \Box \varphi +
    \half{1} g_{F \bar{F}, F \varphi} \partial_\mu \psi \sigma^\mu \bar{\sigma}^\nu
    \partial_\nu \psi \bigr) \psi \sigma^\rho \partial_\rho \bar{\psi}
  \end{split}\\[2ex]
  \begin{split}
\label{eq:pexp4}
    \mathcal{L}^{(4)} &= g_{F \bar{F}} \Box \bar{\varphi} \Box \varphi +
    \inv{4}  g_{F \bar{F}, F \bar{F}} \partial_\mu \psi \sigma^\mu \bar{\sigma}^\nu
    \partial_\nu \psi \partial_\rho \bar{\psi} \bar{\sigma}^\rho \sigma^\lambda
    \partial_\lambda \bar{\psi} \medsp
    &\quad + \half{1} g_{F \bar{F}, \bar{F}} \Box \varphi \partial_\mu \bar{\psi} \bar{\sigma}^\mu \sigma^\nu
    \partial_\nu \bar{\psi} + \half{1} g_{F \bar{F}, F} \Box \bar{\varphi} \partial_\mu \psi \sigma^\mu \bar{\sigma}^\nu
    \partial_\nu \psi
  \end{split}
\end{align}
To get a stable p-expansion up to some given order we have to add terms
including explicit space-time derivatives. The following integral is of main
importance:
\begin{equation}
\label{eq:cadd}
 \mathcal{L}_c = \intd{\diff{^4 \theta}} \sum_k c_k \bar{\Phi} \Phi
 \partial_\mu \Phi \partial^\mu \Box^{k-1} \bar{\Phi} 
\end{equation}
Evaluated with respect to a minimum $\varphi_0$, $F_0$ and $\psi_0 \equiv 0$
we get the following bilinear terms
\begin{align}
  \label{eq:cexp2}
  \mathcal{L}_{c}^{(2)} &= c_1 \bigl( \bar{F} F \partial_\mu \varphi
  \partial^\mu \bar{\varphi} + \bar{\varphi} \varphi \partial_\mu F
  \partial^\mu \bar{F} + ( F \bar{\varphi} \partial_\mu \varphi \partial^\mu
  \bar{F} + \hc \bigr) \medsp
\begin{split}
  \mathcal{L}_{c}^{(2 k)} &= (c_k \bar{F} F - c_{k-1} \bar{\varphi} \varphi) \partial_\mu \varphi
  \partial^\mu \Box^{k-1} \bar{\varphi}\medsp
  &\quad + c_k \bigl( \bar{\varphi} \varphi \partial_\mu F
  \partial^\mu \Box^{k-1}\bar{F} + ( F \bar{\varphi} \partial_\mu \varphi \partial^\mu
  \Box^{k-1} \bar{F} + \hc \bigr)
\end{split} \medsp
  \mathcal{L}_{c}^{(2 k +1)} &= -i c_k \bar{\varphi} \varphi \psi \sigma^\mu
  \partial_\mu \Box^k \bar{\psi}
\end{align}
To cancel off-diagonel terms additional contributions with explicit space-time
derivatives can be introduced. A possible choice, which straightforwardly
generalizes to specific restrictions on the dimensions of the coupling
constants, reads:
\begin{align}
  \begin{split}
\label{eq:dadd}
    \mathcal{L}_{d} &= \intd{\diff{^4 \theta}} d (\Phi_0 \bar{\Phi}_0)
    (\bar{\Phi}_1 \bar{\Phi}_0) \partial_\mu \Phi_0 \partial^\mu \Phi_1 \medsp
    &= d (2 \bar{F} F \bar{\varphi}) (\varphi \partial_\mu F \partial^\mu
    \bar{F} + F \partial_\mu \varphi \partial^\mu \bar{F} ) + \mathcal{O}(p^3)
  \end{split}\medsp
  \begin{split}
\label{eq:eadd}
    \mathcal{L}_e &= \intd{\diff{^4 \theta}} e \bar{\Phi}_1 \Phi_1
    \partial_\mu \bar{\Phi}_0 \partial^\mu \Phi_0 \medsp
    &= e \bar{F} F \partial_\mu \bar{F} \partial^\mu F + \mathcal{O}(p^3) 
  \end{split}\medsp
  \begin{split}
\label{eq:fadd}
    \mathcal{L}_f &= \intd{\diff{^4 \theta}} f \frac{\bar{\Phi}}{\Phi}
    \partial_\mu \Phi \partial^\mu \Phi \medsp
    &= f \bar{F} \bigl(\frac{F}{\varphi^2} \partial_\mu \varphi \partial^\mu
    \varphi + \inv{\varphi} (2 \partial_\mu F \partial^\mu \varphi + \psi \Box
    \psi) \bigr) + \mathcal{O}(p^3)
  \end{split} \medsp
  \begin{split}
\label{eq:gadd}
    \mathcal{L}_g&= \intd{\diff{^4 \theta}} g \bar{\Phi}_1 \Phi_1 \bar{\Phi}_0^2
    \partial_\mu \Phi_0 \partial^\mu \Phi_0 \medsp
    &= 2 g \bar{F}^2 F \bar{\varphi} (2 \partial_\mu F \partial^\mu \varphi +
    \psi \Box \psi) + \mathcal{O}(p^3)
  \end{split}
\end{align}


\bibliography{biblio}
\end{document}